\documentclass[conference,10pt]{IEEEtran}
\IEEEoverridecommandlockouts
\usepackage{cite}
\usepackage{amsmath,amssymb,amsfonts}
\usepackage{algorithm}
\usepackage{algorithmicx}
\usepackage{algpseudocode}
\usepackage{graphicx}
\usepackage{textcomp}
\usepackage{xcolor}
\usepackage{lettrine}
\usepackage{multirow}
\usepackage{caption}
\usepackage{subcaption}
\usepackage{tabularx}
\usepackage{romannum}
\usepackage{array}
\usepackage{cleveref}
\usepackage{url}
\usepackage{geometry}
\crefrangeformat{figure}{Fig.~#3#1--#4#2}
\newcolumntype{P}[1]{>{\centering\arraybackslash}p{#1}}
\newcolumntype{M}[1]{>{\centering\arraybackslash}m{#1}}
\def\BibTeX{{\rm B\kern-.05em{\sc i\kern-.025em b}\kern-.08em
    T\kern-.1667em\lower.7ex\hbox{E}\kern-.125emX}}
\begin{document}
\setlength{\abovedisplayskip}{3pt}
\setlength{\belowdisplayskip}{3pt}
\setlength{\floatsep}{2 pt}
\setlength{\textfloatsep}{2 pt}

\title{Edge-Efficient Deep Learning Models for Automatic Modulation Classification: A Performance Analysis
\vspace*{-0.6em}
}

\author{\IEEEauthorblockN{Nayan Moni Baishya, B. R. Manoj, and Prabin K. Bora}
\IEEEauthorblockA{Department of Electronics \& Electrical Engineering, Indian Institute of Technology Guwahati, Guwahati, Assam, India}
Emails: \text{\tt\{nmb94, manojbr, prabin\} @iitg.ac.in}
\thanks{This work was supported in part by SERB Start-Up Research Grant Scheme, Govt. of India under Grant SRG/2022/001214 and in part by Start-Up Grant of Indian Institute of Technology Guwahati. 

}
\vspace*{-1.75em}
}

\maketitle

\begin{abstract}
The recent advancement in deep learning (DL) for automatic modulation classification (AMC) of wireless signals has encouraged numerous possible applications on resource-constrained edge devices. However, developing optimized DL models suitable for edge applications of wireless communications is yet to be studied in depth. In this work, we perform a thorough investigation of optimized convolutional neural networks (CNNs) developed for AMC using the three most commonly used model optimization techniques: a) pruning, b) quantization, and c) knowledge distillation. Furthermore, we have proposed optimized models with the combinations of these techniques to fuse the complementary optimization benefits. The performances of all the proposed methods are evaluated in terms of sparsity, storage compression for network parameters, and the effect on classification accuracy with a reduction in parameters. The experimental results show that the proposed individual and combined optimization techniques are highly effective for developing models with significantly less complexity while maintaining or even improving classification performance compared to the benchmark CNNs.

\end{abstract}

\begin{IEEEkeywords}
Deep learning, knowledge distillation,  model optimization, pruning, quantization, wireless  classifiers. 
\end{IEEEkeywords}
\vspace*{-0.5em} 
\section{Introduction}
Automatic modulation classification (AMC) of radio frequency (RF) signals is vital to communication systems. The modulation information of signals is essential for further signal demodulation and decoding in practical applications, such as cognitive radio, signal recognition, and spectrum monitoring \cite{nandi, dobre2007survey}. Therefore, AMC is widely applicable in many military and civilian applications. Conventional AMC methods can be categorized into decision theory-based and statistical machine learning-based methods \cite{dobre2007survey}. However, these methods are computationally complex and hard to design. Recently, developing deep learning (DL)-based methods has attracted significant attention in RF modulation classification. DL-based methods have enabled the learning from large-scale RF datasets without any complex feature extraction process. O'Shea et al. \cite{o2016convolutional} proposed the first convolutional neural network (CNN) for modulation recognition, known as VTCNN2. Subsequently, several DL architectures have been applied for AMC, such as ResNet \cite{west2017deep}, InceptionNet \cite{west2017deep}, DenseNet \cite{liu2017deep}, long-short term memory \cite{zhang2020automatic}, etc. The over-parameterized DL architectures with millions of parameters can learn complex representations from extensive training data and  generally require high computing resources for training. However, the drawback of large networks with millions of parameters, as in \cite{o2016convolutional, west2017deep,liu2017deep,zhang2020automatic}, is the difficulty of deploying on resource-constrained edge devices. With the growing number of edge applications and increasing efforts to transfer artificial intelligence (AI) from cloud or centralized applications, efficiently deploying DL models on edge devices has gained significant interest. 
However, edge devices are generally limited in their computational capabilities and are restricted by processing power, memory, and power consumption constraints. 
As AMC is applied at the receiver end, implementing it on the edge devices, such as smartphones, intelligent vehicles, drones, and other IoT devices, can play a significant role. However, current DL-models, as in \cite{west2017deep,liu2017deep,o2016convolutional}, are not memory and compute-optimized to deploy on resource constraint devices.

Much of the work on model optimization \cite{mishra2020survey} has found applications in computer vision. Various popular techniques such as pruning, quantization, and knowledge distillation (KD) are applied to optimize models for smaller model sizes, quicker inference times, and lower power consumption \cite{cheng2017survey}. These commonly used techniques are very effective in optimizing complex and accurate DL models before deployment without making any changes to the network architecture.
However, the performance of these model optimization techniques on DL methods in wireless communication applications, specifically for AMC, needs to be thoroughly investigated. 
Wireless systems often have limited computational resources, are battery-powered, rely on energy harvesting, and require fast real-time data processing. Therefore, DNN models must be optimized for low power consumption, and quicker inference before deploying on edge devices \cite{shi2020communication}.  
The optimization of DNN models for wireless applications consists of a trade-off between complexity, accuracy, storage, and power consumption parameters. By carefully tuning these parameters, it is possible to develop models that meet the requirements of edge devices without affecting the performance of wireless systems and can be deployed in real-time, which is the main scope of this paper. 

The main contributions of this paper are as follows: 
a) in contrast to the complex networks developed for AMC, we propose to develop optimized models using three model optimization techniques, namely, pruning, quantization, and KD, and evaluate their performances; 
b) we establish the effectiveness of combining optimization strategies to merge the benefits of individual methods. Such methods can be beneficial to develop smaller and optimized models with similar performance to large, complex, but highly accurate models; 
c) the proposed optimal models offer significant benefits, such as high network sparsity, high compression rate, and reduced parameters for efficient deployment on resource-limited devices. Despite these advantages, the accuracy of our models is comparable to the original complex models and even outperforms them when using KD as part of the optimization process.


\section{Network Architectures and Model Optimization Methods}
In this section, we describe on three individual model optimization methods: a) network pruning, b) model quantization, and c) knowledge distillation, along with two combined strategies for optimizing CNN architectures developed for AMC
\vspace*{-0.3em}
\subsection{Network Architectures}
We have considered three CNN-based benchmark architectures, namely, VTCNN2 \cite{o2016convolutional}, ResNet \cite{west2017deep}, and InceptionNet \cite{west2017deep}, for applying the proposed model optimization techniques. The primary motivation for selecting these models as the benchmark is the complexity of the networks due to a large number of parameters. We have chosen these large networks to demonstrate the effectiveness of different optimization techniques and, in general, are applicable to any other model with high complexity. The VTCNN2 comprises about 2.83M parameters, and the ResNet and the InceptionNet architectures have about 3.45M and 10.14M parameters, respectively. It is observed in all three considered CNN models that the first fully connected (FC) layer contains the highest number of parameters. For VTCNN2, this layer has 2.70M of 2.83M total parameters, ResNet has 3.37M out of 3.45M total parameters, and InceptionNet has 10.06M out of 10.14M total parameters in their respective first FC layers. Based on this observation, we choose to optimize this parameter-heavy layer, which, in turn, will optimize the overall models.
\vspace*{-0.2em}
\subsection{Network Pruning}
Network pruning is a powerful model optimization technique that can improve the efficiency of a DNN by removing certain connections from the network while preserving its accuracy. This is based on the fact that DNNs are highly over-parametrized, and many parameters do not contribute significantly to the network's performance \cite{mishra2020survey}. Pruning these less important parameters can reduce overfitting and improve the generalization of a DNN. A pruned network also has a smaller memory footprint, requires fewer computations, and can be deployed on edge devices with limited resources. 
In this paper, we have considered an unstructured pruning technique called the \textit{Net-trim} (NT) algorithm \cite{aghasi2017net}. The main idea behind the NT algorithm is to maximize the sparsity of weights for each layer of a trained neural network (NN) such that the post-pruning output responses remain consistent with the initial output responses. It can be formulated as a constrained optimization problem for layer $l$ as given by \cite{aghasi2017net} 
\begin{eqnarray} \label{eq:1}
\small
   && \hat{\mathbf{W}}_{l}= \underset{\mathbf{U}_{l}}{\arg\min} \|\mathbf{U}_{l}\|_1 \\ &&
    \text{s.t.} \quad  \|\hat{\mathbf{Y}}_l- \mathbf{Y}_{l}  \|_{F} \leq \epsilon\, ,   \nonumber
\end{eqnarray}
where $\hat{\mathbf{Y}}_l= \max(\mathbf{U}_{l}^\top \mathbf{X}_{l}, 0)$ with $(\cdot)^\top$ denotes transpose, $\|\cdot\|_F$ is the Frobenius norm, and $\epsilon >0$ is the threshold. In (\ref{eq:1}), $\hat{\mathbf{W}}_{l}$ is the sparse weight matrix of layer $l$, $\mathbf{X}_{l} \in \mathbb{R}^{N \times P}$ is the stacked input to the layer with $N$ being the number of samples and $P$ as the dimension of each sample, $\mathbf{Y}_{l}$ is the output activation of the layer before pruning, $\hat{\mathbf{Y}}_l$ is the output activation of the layer during optimization for the intermediate weight matrix $\mathbf{U}_{l}$. 
Here, the non-linear activation function ReLU is considered for the layer. The optimization problem in (\ref{eq:1}) is solved using the alternating direction method of multipliers (ADMM) technique \cite{aghasi2017net}. 
The implementation of the NT algorithm on the weight matrix of the first FC layer of a DL-based AMC network is presented in Algorithm \ref{alg:net_trim}. 
In the algorithm, the $\textsc{TRIM}$ procedure depicts the iterative solution of the optimization problem in (\ref{eq:1}) to obtain the sparse weight matrix $\hat{\mathbf{W}}_{l}$ and the parameter $\epsilon$ controls the sparsity of the weights. The input data and output activations of the $l^\text{th}$ network layer are given by $\mathbf{Y}_{l-1}$ and $\mathbf{Y}_{l}$, respectively. $L$ denotes the total number of layers, and $k$ denotes the index of the first FC layer of the CNN architectures that are considered in our work.
\begin{algorithm}[t!]
\small
\caption{Net-trim }
\label{alg:net_trim}
\textbf{Input:}  Data matrix $\mathbf{X} \in \mathbb{R}^{N \times P}$, \\
$\text{normalized weight matrices} \, \mathbf{W}_1,\mathbf{W}_2, \dots, \mathbf{W}_L, \\\text{index } \, \, k \,\, \text{and} \,\, \epsilon > 0$
\begin{algorithmic}[1]
\State $\mathbf{Y}_{0} \gets \mathbf{X} \text{\qquad \qquad \qquad \qquad \quad \,\, // Input data}$ 
\For{$l=1$ to $L$}    
      \State $\mathbf{Y}_{l} \gets \max(\mathbf{W}_l^\top \mathbf{Y}_{l-1}, \mathbf{0})$\,\,\,\,\, // Generating pre-pruning layer outputs
    \EndFor
\State $\hat{\mathbf{W}}_{k} \gets \textsc{TRIM}(\mathbf{Y}_{k-1}, \mathbf{Y}_{k}, \mathbf{0}, \epsilon)$ // Apply $\textsc{TRIM}$ on FC layer $k$
\end{algorithmic}
\textbf{Ouput:} Pruned weight matrix $\hat{\mathbf{W}}_{k}$
\end{algorithm}
\subsection{Model Quantization}
Model quantization involves compressing the weights of different CNN layers to  lower precision representations than the original high precision. For example, in an FC layer, the weights can be stored using 8-bit integers instead of using 64-bit floating-point numbers. The quantization of weights significantly reduces the storage requirements for deployment on edge devices with limited memory. It was demonstrated in \cite{denil2013predicting} that a small subset ($5\%$) of the parameters can accurately predict the layer weights of a DNN due to the redundancies present in NN parameter space. With this motivation, we have considered vector quantization, which is successfully used in signal processing applications to exploit the redundancies in a high-dimensional data space. Specifically, we have implemented the \textit{product quantization} (PQ) algorithm \cite{jegou2010product} to compress the weight matrices of the first FC layers of the three complex networks: VTCNN2, ResNet, and InceptionNet.A mathematical representation of the implementation of the PQ algorithm is presented in Algorithm \ref{alg:product_quantization}, where $\mathbf{W} \in \mathbb{R}^{M \times N}$ denotes the weight matrix of an FC layer and it is divided into $P$ subspaces column-wise, each of dimension $M \times \left(\frac{N}{P}\right)$. For each subspace, \textit{$K$-means clustering} is performed to obtain the $K_s$ centroids, and all the sub-vectors are assigned to one of the centroids. The original $N$-dimensional vector (row) can now be represented with $P$-dimensional vector, which is indexed with an integer-valued PQ code. Each value in the code is the cluster index of a sub-vector in the respective subspace. During inference, the quantized sub-vectors are used to reconstruct an approximation of the original high-dimensional weight vector. The compression rate for this method is
\begin{equation}
\small
    \mathit{C_Q}=\frac{bMN}{bK_{s}N + \log_2(K_{s})MP} \, ,
\end{equation}
where $b$ is the number of bits used to represent the original weights and $\mathit{C_Q}$ signifies the ratio of the number of bits required to store the original weight matrix to the number of bits required to store the PQ codes along with the codebook.
\begin{algorithm}[t!]
\small
\caption{Product quantization}\label{alg:product_quantization}
\textbf{Input:}  Weight matrix $\mathbf{W}\in\mathbb{R}^{M\times N}$,  $P$, $K_s$, 
$M>K_s$, $N$ is divisible by $P$
\begin{algorithmic}[1]
\State Partition $\mathbf{W}$ column-wise into $P$ disjoint submatrices: $\mathbf{W}=[\mathbf{W}^{1},\mathbf{W}^{2}, \cdot\cdot\cdot,\mathbf{W}^{P}]$, where $\mathbf{W}^{i} \in \mathbb{R}^{M \times d}, d=\frac{N}{P}$ 
\For{\textbf{each} $\mathbf{W}^{i}, i=1 \dots P$} 
\State \textit{K-means clustering}($\mathbf{W}^{i}$, $K_s$)
\State \textbf{return:} The final cluster centroids $\{\mathbf{c}^{i}_j\}^{K_s}_{j=1}$, $\mathbf{c}^{i}_j \in \mathbb{R}^d$
\State Concatenate all the centroids row-wise to form the sub-codebook $\mathbf{C}^{i} \in \mathbb{R}^{K_s \times d}$ 
\State Assign each subvector, $\mathbf{w}^{i}_{z} \in \mathbf{W}^{i} $ to its closest centroid: $\mathbf{c}^i_{j} \gets \operatorname{argmin}_{j} d(\mathbf{w}^{i}_{z}, \mathbf{c}^{i}_j)$ 
\State Store the cluster index for $\mathbf{w}^{i}_{z}$ as $\overline{w}^{i}_{z}$   
\EndFor
\State The PQ code for a $N$-dimensional vector $\mathbf{w}_{z} \in \mathbf{W}$ is $\overline{\mathbf{w}}_z=[\overline{w}^{1}_{z},\overline{w}^{2}_{z}, \dots,\overline{w}^{P}_{z}] \in \{ 1,2, \dots,K_s \}^P$
\State The overall codebook is $\mathbf{C}=[\mathbf{C}^{1}, \mathbf{C}^{2}, \dots, \mathbf{C}^{P}]$
\end{algorithmic}
\textbf{Output:} The PQ codes $\forall  \mathbf{w}_z \in \mathbf{W}$: $\{ \overline{\mathbf{w}}_z\}^{M}_{z=1}$ and the codebook $\mathbf{C} \in \mathbb{R}^{P \times K_s \times d}$
\end{algorithm}
\vspace*{-0.3em}
\subsection{Knowledge Distillation (KD)}
KD is a technique to transfer knowledge from a more complex network (teacher model) to a smaller, lightweight network (student model). The teacher model provides additional guidance and regularization, which can help the student model avoid overfitting and improve its generalization performance. The smaller student model is more computationally efficient, capable of faster inference, and easier to deploy on devices with limited resources. Hinton et al. \cite{hinton2015distilling} introduced the concept of vanilla KD framework, which is presented in  Algorithm \ref{alg:knowledge_distillation} and demonstrated its effectiveness in reducing the size of DNNs while maintaining high accuracy.
In the algorithm, the pre-trained teacher model is denoted by ${f}^{t}_{\mathbf{V}}$ and the student model is denoted by ${f}^{s}_{\mathbf{W}}$, where $\mathbf{V}$ and $\mathbf{W}$ are the respective trainable model parameters. $\textsc{BATCH}(\cdot)$ and $\textsc{TRAIN}(\cdot)$ represent the sampling of random batches from the training dataset and the training of a model batch-wise, respectively. 
the temperature parameter $T$ in the $\text{\textit{Softmax}}_T(\cdot)$ activation function regulates the softness of the output probability distributions. During the distillation process, a high value of $T$ is used, but after the distilled model is trained,  $T$ is set to $1$. The distillation loss, i.e., the distance between the output probability distributions $\mathbf{\hat{Y}}^{t}_{B}$ and $\mathbf{\hat{Y}}^{s}_{B}$ for teacher and student models respectively, is minimized using the Kullback-Leibler (KL) divergence, $\textit{KL}_{Divergence} (\cdot) $. The overall loss function is the weighted sum of distillation loss and cross-entropy loss for the student model with $\alpha $ being the weight of the distillation loss. 

\vspace*{-0.5em}
\begin{algorithm}[t!]
\small
    \caption{Vanilla knowledge distillation}
    \label{alg:knowledge_distillation}
\textbf{Input:} Teacher model ${f}^{t}_{\mathbf{V}}$, student model ${f}^{s}_{\mathbf{W}}$, training dataset $\mathcal{D}$, batch size $B$, temperature $T$, weight of distillation loss $\alpha$\
\begin{algorithmic}[1]
\State  $\mathbf{\hat{V}} \gets \textsc{TRAIN}( {f}^{t}_{\mathbf{V}}, \mathcal{D}, \mathcal{L}_{c}, \text{\textit{Softmax}}_T $) \, \, \, \, // $\mathbf{\hat{V}}$ are the trained parameters
\State Initialize $\mathbf{W}$ randomly
\Repeat
\State  $(\mathbf{X}_B,\mathbf{Y}_B) \gets \textsc{Batch}(\mathcal{D})$
\State  $\mathbf{\hat{Y}}^{t}_B \gets \text{\textit{Softmax}}_T({f}^{t}_{\mathbf{V}}(\mathbf{X}_B))$ 
\State $\mathbf{\hat{Y}}^{s}_B \gets \text{\textit{Softmax}}_T({f}^{s}_{\mathbf{W}}(\mathbf{X}_B))$ 
\State $  \mathcal{L}_{d} = \textit{KL}_{Divergence}(\mathbf{\hat{Y}}^{t}_B, \mathbf{\hat{Y}}^{s}_B)$ 
\State $\mathcal{L}_{c}=\text{\textit{Cross-entropy}}(\mathbf{\hat{Y}}^{s}_B, \mathbf{Y}_B)$ 
\State Compute the total loss as: $\mathcal{L} = \alpha\cdot\mathcal{L}_{d} + (1-\alpha)\cdot\mathcal{L}_{c}$
\State Update: $\mathbf{W} \gets \mathbf{W}-\nabla_\mathbf{W}\mathcal{L}$
\Until{convergence}
\end{algorithmic}
\textbf{Output:} Distilled student model ${f}^{s}_\mathbf{W}$
\end{algorithm}
\subsection{Combined Optimization Strategies for AMC}
\vspace*{-0.1em}
All three techniques discussed above have unique and complementary advantages for DL model optimization. Moreover, these techniques do not significantly affect the classification performance, as shown in Section \Romannum{3}, which is most important for AMC model optimization. Inspired by this, we propose two strategies to fuse the benefits of the techniques under consideration, which are named the \textit{Distilled pruning} (DP) and the  \textit{Distilled quantization} (DQ) methods. These methods are outlined below: 
\paragraph{Distilled pruning} This method implements KD as the first step of optimization, followed by applying the NT algorithm on the FC layer weights of the distilled student model. The transfer of knowledge through KD can help to achieve better model performance with a small model, and the NT method is highly effective in sparsifying the model without affecting the performance. Through this method, we can obtain an optimized model with less parameter count and high model sparsity, translating to better storage and computational efficiency during edge inference.

\paragraph{Distilled quantization} Similar to DP, the DQ method also performs KD on a small student model, followed by implementing PQ on the FC layer weights to learn a codebook to store the weights. The final optimized model is benefitted from the better feature representations learned from KD as well as a reduction in precision and the number of weights to be stored in a memory-constrained device. 

\vspace*{-0.5em}
\section{Results and discussions}
\begin{figure*}[!t]
     \centering
     \begin{subfigure}[h]{0.32\textwidth}
         \centering
         \centerline{\includegraphics[width=5.4cm, height=4cm]{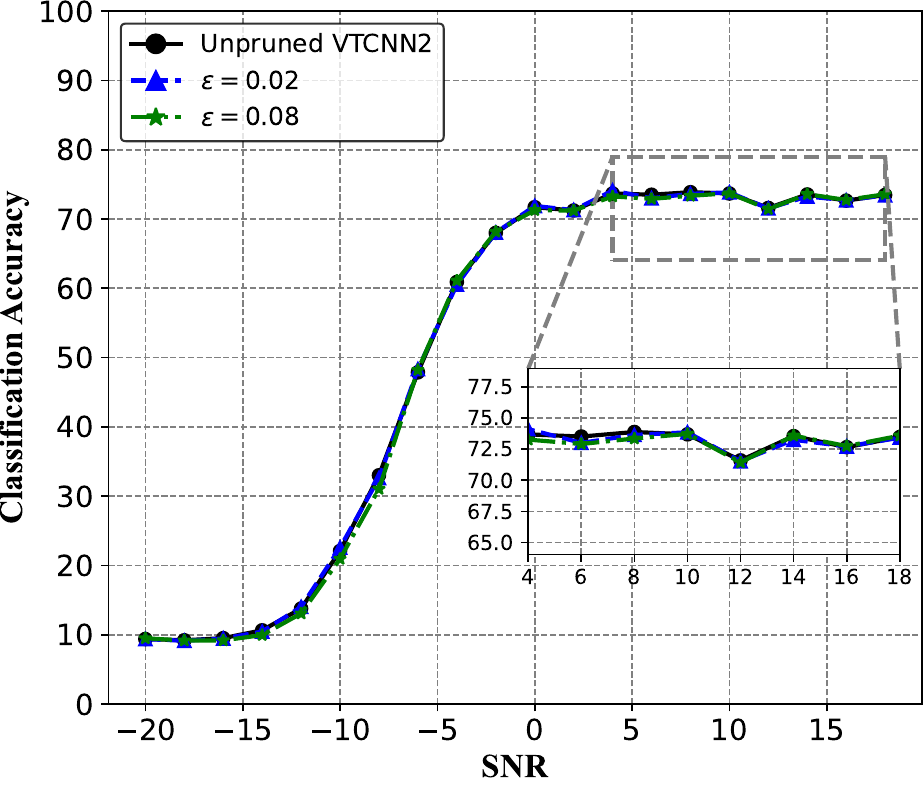}}
         \caption{VTCNN2}
         \vspace*{-0.3em}
         \label{fig:trim-vtcnn}
     \end{subfigure}
     \begin{subfigure}[h]{0.32\textwidth}
         \centering
         \centerline{\includegraphics[width=5.4cm, height=4cm]{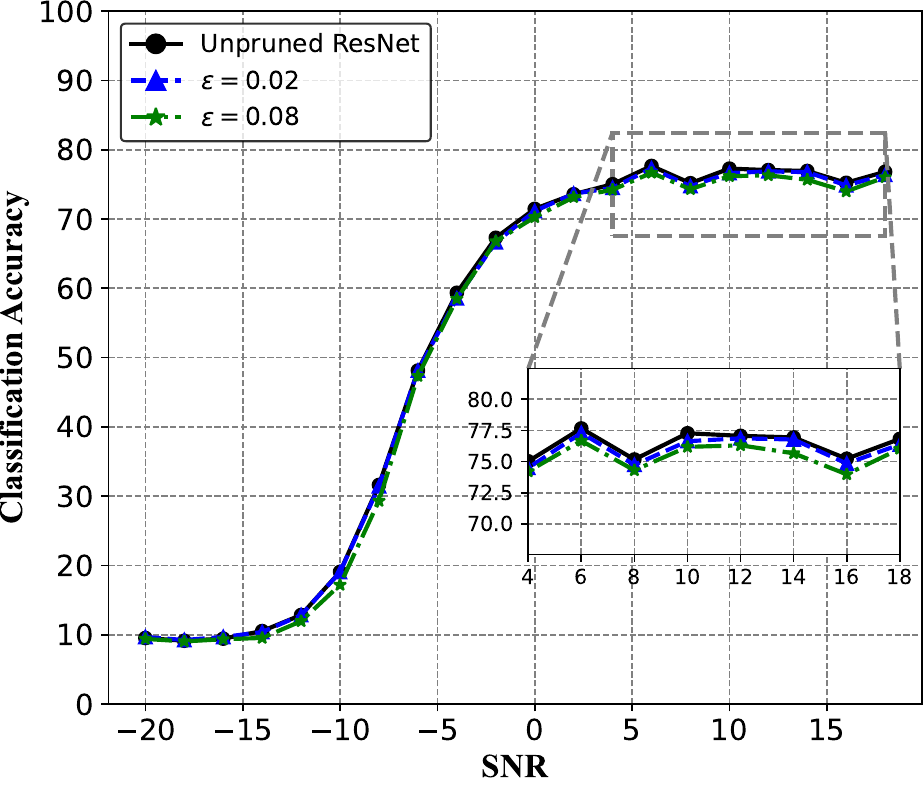}}
         \caption{ResNet}
          \vspace*{-0.3em}
         \label{fig:trim-res}
     \end{subfigure}
     \begin{subfigure}[h]{0.32\textwidth}
         \centering
         \centerline{\includegraphics[width=5.4cm, height=4cm]{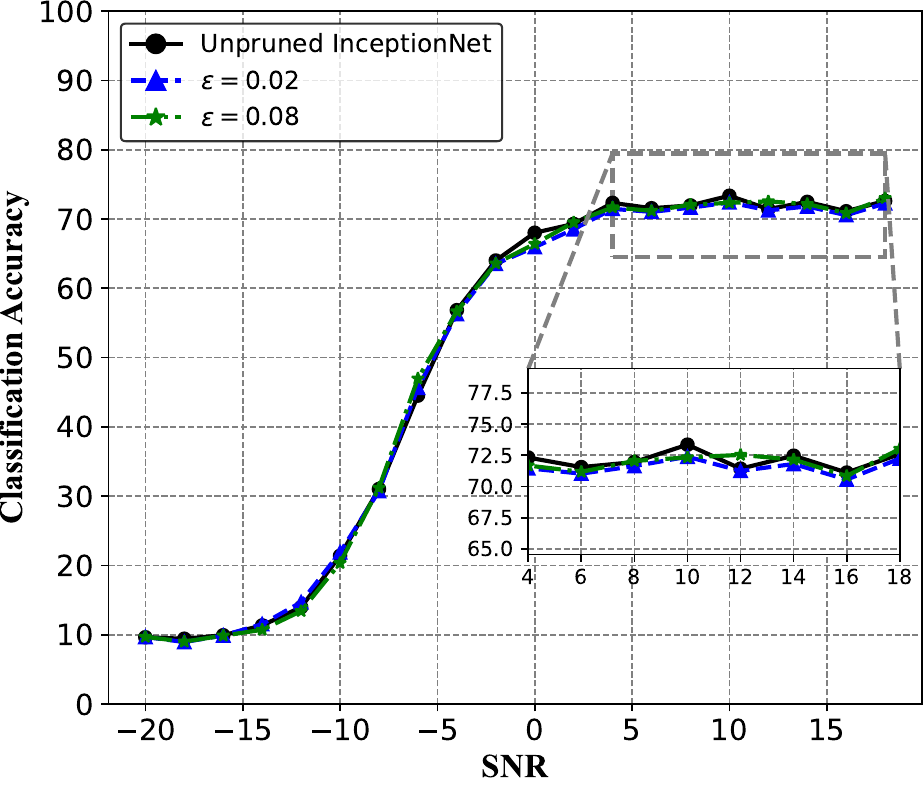}}
         \caption{InceptionNet}
         \vspace*{-0.5em}
         \label{fig:trim-inc}
     \end{subfigure}
        \caption{\small Performance of network pruning method (NT) for different values of $\epsilon$. } \vspace*{-0.5em}
        \label{fig:net-trim}
\end{figure*}

In this section, we will discuss the results of the proposed models developed using individual and combined optimization techniques. For all the experiments, we consider the RadioML2016.10A AMC dataset \cite{vtcnn2}, consisting of the $11$ most widely used modulation schemes. For each modulation scheme, there are 20,000 samples belonging to $20$ different signal-to-noise ratios (SNRs), varying from $-20$dB to $18$dB with a step of $2$dB. Each RF signal in the dataset is of dimension $2 \times 128 $ with both real and imaginary parts. 
We have considered the RadioML2016.10A dataset as it is publicly available and for reproducibility of the results. All three network architectures are first pre-trained on this RF dataset using $50\%$ of data as the training dataset ($\mathrm{D_{Train}}$) and the remaining $50\%$ for testing ($\mathrm{D_{Test}}$) \cite{vtcnn2}. Evaluation on a larger test set eliminates training data bias on the results of the optimized models and will reinforce generalization on real-world test samples. The benchmark classification accuracy versus SNR plots for all three networks is shown in \cref{fig:trim-vtcnn,fig:trim-res,fig:trim-inc} and are used to compare the performances of the proposed optimized models.
\begin{figure*}[!]
     \centering
     \begin{subfigure}[h]{0.32\textwidth}
         \centering
         \centerline{\includegraphics[width=5.4cm, height=4cm]{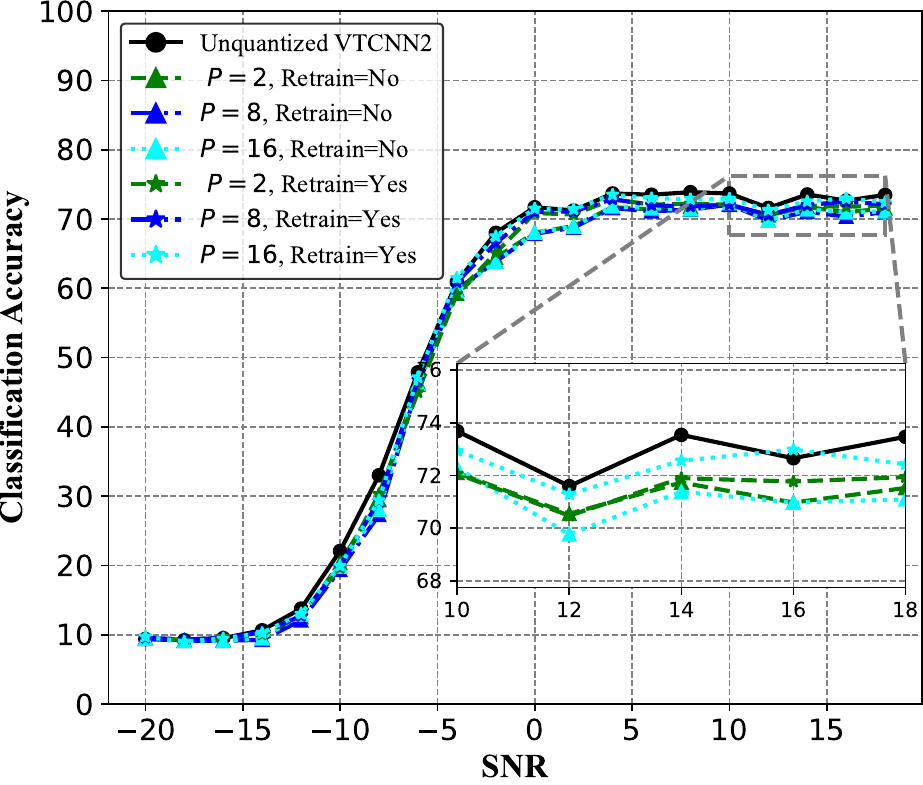}}
         \caption{VTCNN2}
         \vspace*{-0.3em}
         \label{fig:quant_vtcnn}
     \end{subfigure}
     \begin{subfigure}[h]{0.32\textwidth}
         \centering
         \centerline{\includegraphics[width=5.4cm, height=4cm]{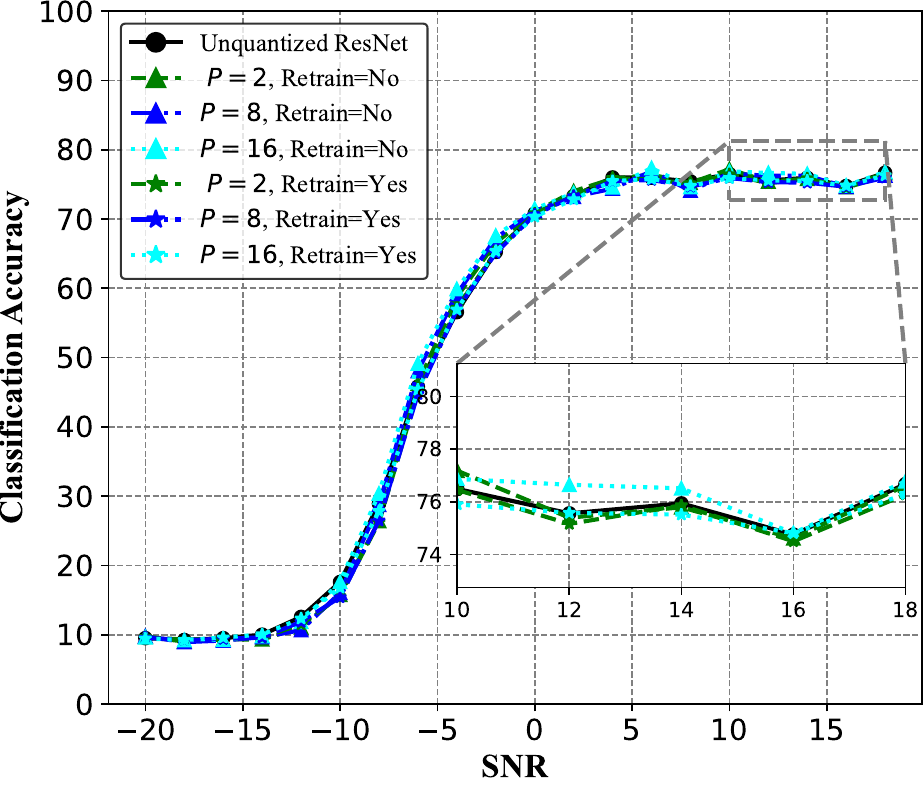}}
         \caption{ResNet}
         \vspace*{-0.35em}
         \label{fig:quant_resnet}
     \end{subfigure}
     \begin{subfigure}[h]{0.32\textwidth}
         \centering
         \centerline{\includegraphics[width=5.4 cm, height=4cm]{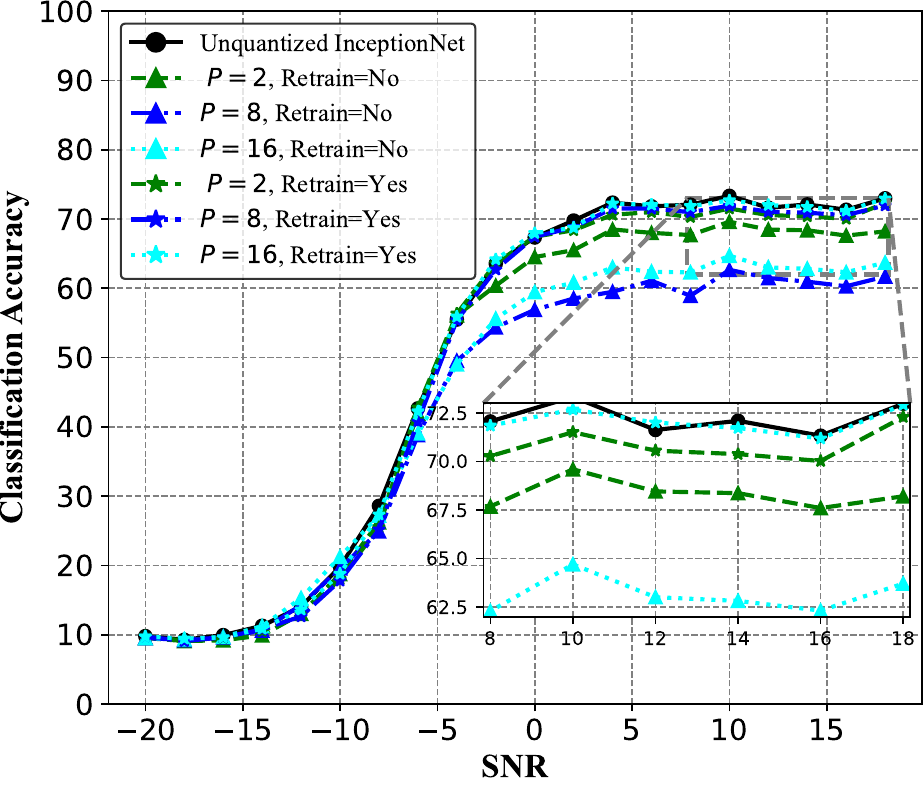}}
         \caption{InceptionNet}
         \vspace*{-0.35em}
         \label{fig:quant_incep}
     \end{subfigure}
        \caption{\small Performance of model quantization method (PQ) with and without retraining for different values of $P$.}
         \vspace*{-0.35em}
        \label{fig:quantization}
\end{figure*}
\begin{figure*}[t]
     \centering
     \begin{subfigure}[h]{0.32\textwidth}
         \centering
         \centerline{\includegraphics[width=5.4cm, height=4cm]{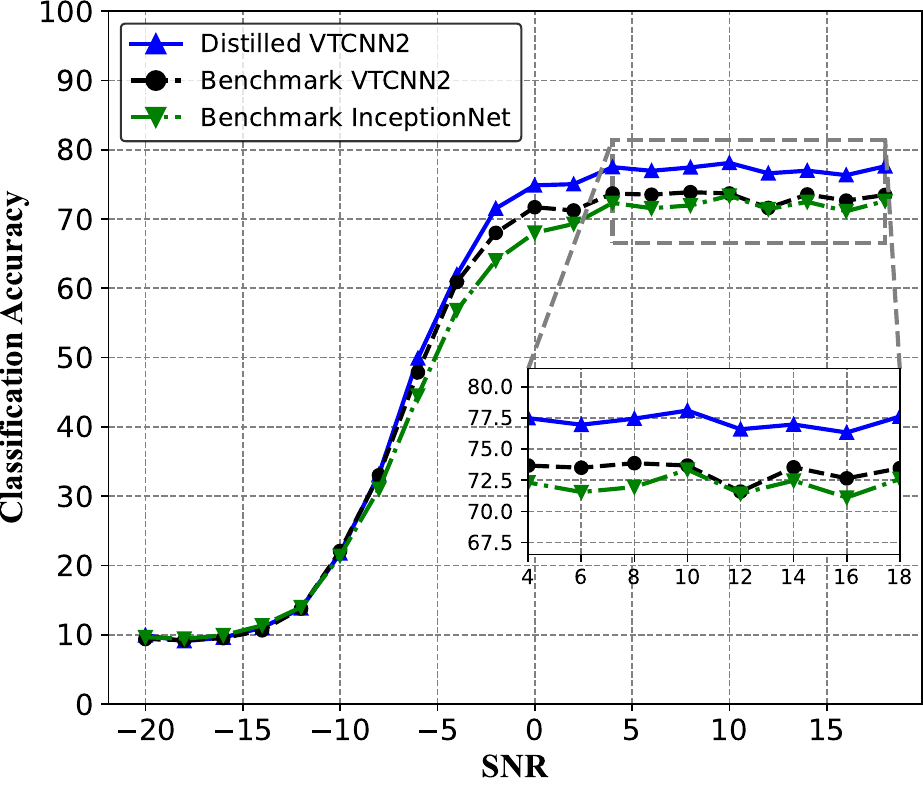}}
         \caption{ Case \Romannum{1} \vspace*{-0.35em}}
         \label{fig: inc-vt}
     \end{subfigure}
     \begin{subfigure}[h]{0.32\textwidth}
         \centering
         \centerline{\includegraphics[width=5.4cm, height=4cm]{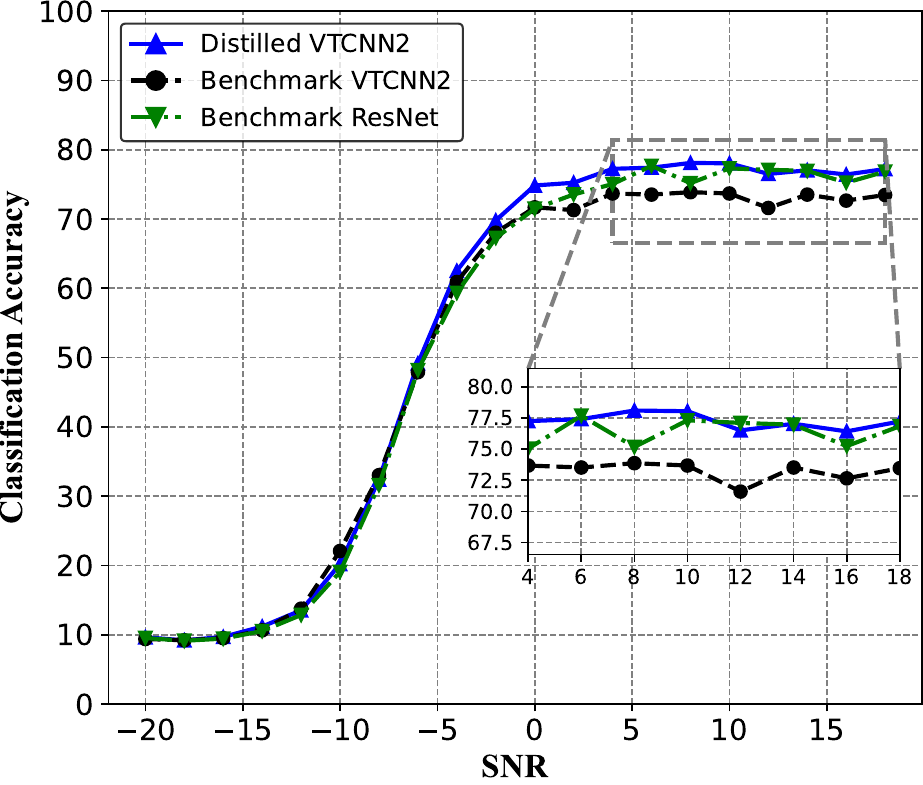}}
         \caption{Case \Romannum{2}\vspace*{-0.35em}}
         \label{fig:res-vt}
     \end{subfigure}
     \begin{subfigure}[h]{0.32\textwidth}
         \centering
         \centerline{\includegraphics[width=5.4cm, height=4cm]{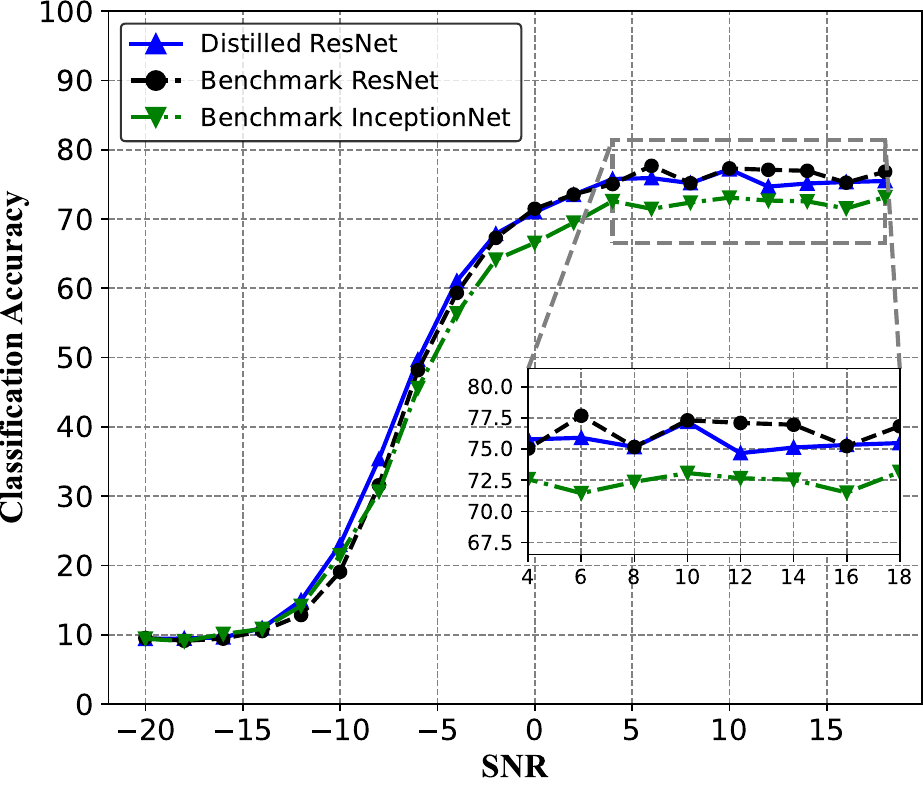}}
         \caption{ Case \Romannum{3}\vspace*{-0.35em}}
         \label{fig:inc-res}
     \end{subfigure}
        \caption{\small Performance of the KD method for all three combinations as given in Table      \ref{tab:kd_cases} with fixed temperature $T =10$.}
         \vspace*{-1.2em}
        \label{fig:KD}
\end{figure*}
\begin{table}[!t]
    \centering
    \caption{\small Pruning efficiency ($p_e$) achieved for $\epsilon=0.08$.}
    \renewcommand{\arraystretch}{1}
    \setlength{\tabcolsep}{5pt}
    
    \begin{tabular}{|M{1.4cm}|M{0.8cm}|M{0.8cm}|M{1cm}|M{1cm}|}
    \hline
    Network & $n_T$ & $n_b$ & $n_a$ & $p_e$\\
    \hline
    VTCNN2 & 2.70M & 2.69M & 94.47K & 96.5\% \\
    \hline
    ResNet & 3.38M & 3.37M  & 63.94K & 98.1\%\\
    \hline
    InceptionNet & 10.11M & 10.07M & 106.48K & 98.94\%\\
    \hline
    \end{tabular}
        \label{tab:np_pe}
\end{table}
\vspace*{-0.3em}
\subsection{Network Pruning}
\vspace*{-0.5em}

To develop the optimized models, the NT pruning algorithm is applied to the parameters of the first FC layers of the three pre-trained complex models.
We have experimented with the parameter $\epsilon = \{ 0.02, 0.08\}$, which controls the amount of sparsity introduced in the model. The input data matrix $\mathbf{X}$ in Algorithm \ref{alg:net_trim} is created from 20,000 randomly selected RF signals from $\mathrm{D_{Train}}$. The performance of the pruning method is evaluated using (a) the \textit{pruning efficiency} ($p_e$) and (b) the classification accuracy as a function of SNRs. The $p_e$ can be defined as
\begin{equation}
    p_e=1-\frac{n_a}{n_T} \,  ,
\end{equation}
where $n_T$ is the total number of parameters of the FC layer, and $n_a$ is the number of non-zero parameters after pruning. Thus, $p_e$ measures the sparsity after the pruning process, and a high value of $p_e$ indicates high sparsity. In our work, we could obtain $p_e \geq 90\% \,\,\, \forall \, \epsilon$, with the highest is achieved for $\epsilon=0.08$. Table \ref{tab:np_pe} shows the $p_e$ for all three networks when $\epsilon=0.08$, where $n_b$ (and $n_a$) is the number of non-zero parameters of the layer before (and after) pruning is applied. The $p_e$ values for VTCNN2, ResNet, and InceptionNet are $96.5\%, 98.1\%, 98.94\%$ respectively for $\epsilon=0.08$. This establishes that only a small fraction of parameters contribute to the modulation classification task and thus significant optimization can be achieved. Also, Fig. \ref{fig:net-trim} compares the accuracy versus SNR curves for unpruned and pruned networks. As seen from \cref{fig:trim-vtcnn,fig:trim-res,fig:trim-inc}, the classification performance always remains comparable to the benchmark, irrespective of the very high sparsity achieved in the optimized models. 
In conclusion, developing optimized models for AMC using the NT algorithm can achieve high sparsity with comparable classification performance to the benchmark. Sparse DNNs can offer compressed storage requirements by only storing the non-zero elements and their indices. The computation complexity during inference is also reduced as the number of operations decreases. However, most of today's popular CPU and GPU architectures and DL frameworks are mainly optimized for dense computations and lack the functions to benefit from the sparsity in DNNs \cite{yu2017scalpel}. With growing research on developing sparse DNNs, the development of hardware accelerators optimized for sparse matrix computations for edge devices is advancing \cite{zhang2020snap}. Such developments will facilitate the applications of sparse DNNs in wireless systems at the edge. 
\vspace*{-0.5em}
\begin{table}[!t]
    \centering
    \caption{\small Compression rates ($C_Q$) for different values of $P$. \vspace*{-0.3em}}
    \renewcommand{\arraystretch}{1}
    \setlength{\tabcolsep}{5pt}
    
    \begin{tabular}{|M{1.4cm}|M{0.8cm}|M{0.8cm}|M{1cm}|M{1cm}| }
    \hline
        \centering Network & ${P=2}$ & ${P=8}$& ${P=16}$  \\
        \hline
         VTCNN2 & 39.65 & 35.52 & 31.2 \\
         \hline
         ResNet & 49.56 & 42.92 & 36.76 \\
         \hline
         InceptionNet & 133.20 & 95.8 & 69.76\\
         \hline
    \end{tabular}
    \label{tab:pq_cm}
\end{table}
\subsection{Model Quantization}
Model quantization using the PQ algorithm is also applied to the first FC layers of all three complex networks. As in Algorithm \ref{alg:product_quantization}, the high-dimensional weight matrix of the FC layer is partitioned into $P$ sub-matrices, and we have experimented with $P= \{2, 8, 16\}$. The number of cluster centroids computed for each sub-matrix is $K_s =256 $ and thus, the PQ codes can be represented with only 8 bits. Table \ref{tab:pq_cm} compares the compression rates ($C_Q$) achieved for the optimized models with different values of $P$. The highest compression rate is achieved for $P=2$. High compression leads to a significant reduction in the total number of bytes required to store the layer parameters as compared to the original $64$-bit float representation. For example, the storage requirement of the FC layer weights of the quantized InceptionNet is reduced by a factor of $133.20$ for $P=2$. Similarly, the PQ code-based storage requirement for ResNet after quantization is reduced by a factor of $49.56$ than the original weight matrix when $P=2$. 

We have compared the classification performance of the quantized networks with the benchmark results, as shown in Fig \ref{fig:quantization}. From Fig. \ref{fig:quant_vtcnn} and Fig. \ref{fig:quant_resnet}, it is observed that the PQ-based quantization of the layer weights results in a minimal loss in accuracy (within $1\%$ of the benchmark) for VTCNN2 and ResNet for all $P$. Specifically, for $P=2$, i.e., when the highest compression is achieved, the accuracies are comparable to the benchmark for all three optimized networks. In the case of InceptionNet, the accuracies are reduced when $P=8,16$. It is observed that the accuracy can be improved with retraining for $20$ epochs using only 10\% of the original training dataset. For retraining, $\mathbf{\hat{W}}$ for the FC layer is kept non-trainable so that the weights remain consistent with the centroids in the codebook and also reduce the number of trainable parameters. 
As seen in Fig. \ref{fig:quant_incep}, the accuracies of the quantized InceptionNet model for  $P=8\, \text{and}\, 16$ can be significantly recovered and become comparable to the benchmark performance with a computationally efficient retraining process.
Thus, it can be established that the PQ algorithm is an effective method for developing optimized models from any large and complex AMC network, with consistent storage benefits for the edge while providing performance comparable to the unquantized networks.

    
\begin{table}[!t]
    \centering
    \caption{\small Combinations of student and teacher models and respective change in parameters and accuracies. \vspace*{-0.3em}}
    \renewcommand{\arraystretch}{1}
    \vspace*{-0.3em}
    \setlength{\tabcolsep}{5pt}
    \begin{tabular}{|M{1cm}|M{1.2cm}|M{1.4cm}|M{1.1cm}|M{1cm}|M{1cm}|}
    \hline
        \centering Cases & Student & Teacher & \shortstack{Reduction\\ ratio} & \multicolumn{2}{|M{1.3cm}|}{\shortstack{Accuracy \\ compared to \\ benchmark}} \\ 
    \hline
         \Romannum{1} & VTCNN2 & InceptionNet &  0.28 & \multicolumn{2}{|M{1.3cm}|}{Improved} \\ 
    \hline
         \Romannum{2} & VTCNN2 & ResNet &  0.82 & \multicolumn{2}{|M{1.3cm}|}{Improved} \\ 
    \hline
         \Romannum{3} & ResNet & InceptionNet &  0.34 & \multicolumn{2}{|M{1.3cm}|}{Comparable} \\
    \hline
    \end{tabular}
    \label{tab:kd_cases}
\end{table}
\subsection{Knowledge Distillation (KD)}
To evaluate the efficacy of the KD method, we have formulated three scenarios of student and teacher models with the student model having fewer parameters compared to the teacher model.
Table \ref{tab:kd_cases} shows the parameter \textit{reduction ratios} for all three cases, which is the ratio of the parameter count of the student to that of the teacher. In case \Romannum{1}, the number of parameters of the VTCNN2 student model is $0.28 \times$ the parameters of the InceptionNet teacher model, which is the smallest among the three cases. The distilled student model is trained on $\mathrm{D_{Train}}$ and the temperature ($T$) for the softmax function is assumed to be $10$. A comparison of the classification performances of the distilled student models with the original benchmark student and teacher models is shown in Fig. \ref{fig:KD} for all three cases. It can be observed from Figs. \ref{fig: inc-vt} and  \ref{fig:res-vt} that the accuracy of the distilled VTCNN2 has improved compared to the benchmark VTCNN2,  with InceptionNet and ResNet being teacher models, respectively. This can benefit model optimization as VTCNN2 is the smallest of all three models, with better accuracy. In case \Romannum{3}, when ResNet is used as the student model, the distillation process only results in comparable accuracy with the benchmark, as shown in Fig. \ref{fig:inc-res}. Thus, it can be said that the choice of the student and teacher architectures plays a vital role in performance improvement. We also observed that the parameter $T$ does not effect the classification performance significantly.
As observed in Case \Romannum{1} and \Romannum{2} of Table \ref{tab:kd_cases}, when the goal is to design a smaller optimized model with a low-complexity architecture, such as VTCNN2, the KD can reduce the effective parameter count while increasing the accuracy of such a model.
\subsection{Combined methods: DP and DQ}
\begin{figure}[!]
    \centering
    \includegraphics[width=2.3in, height=1.6in]{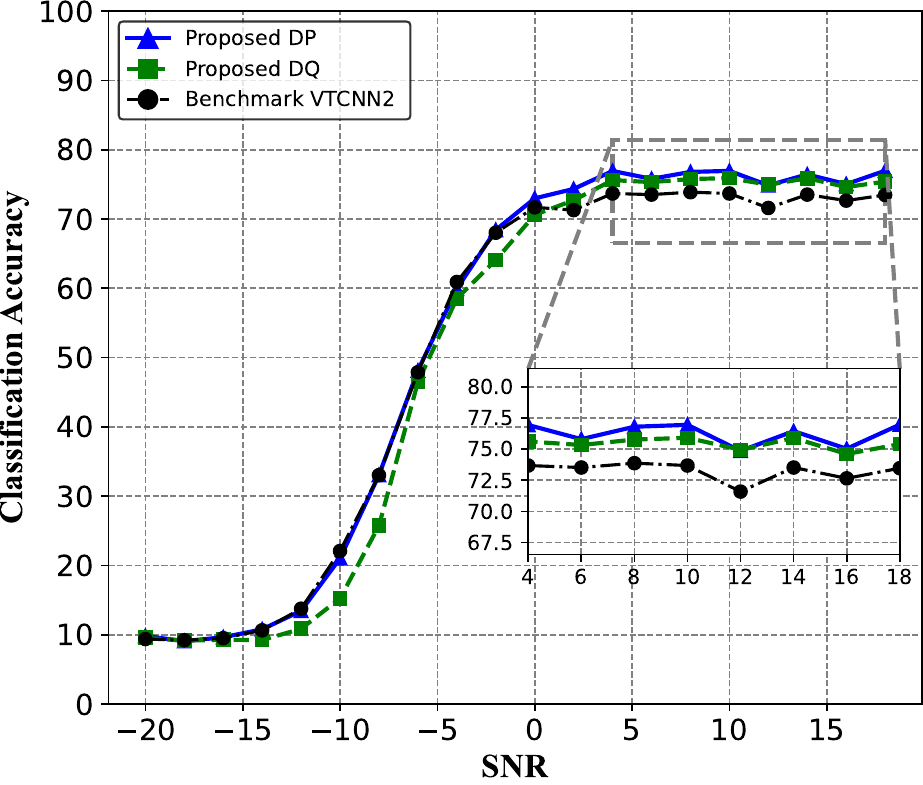}
   \caption{ {\small Comparison of performance of the proposed combined optimization strategies with benchmark VTCNN2.\vspace*{-0.5em}}
    }
    \label{fig:dp-dq}
\end{figure}

We now evaluate the performance of the proposed two combined optimization strategies: the DP and the DQ methods. 
The primary focus is to investigate the effect on classification
performance while attempting to merge the benefits of two
optimization techniques simultaneously. Optimizing a model with less complexity while improving
classification performance is desirable for edge applications.
Therefore, the VTCNN2 model is optimized using the two
combined methods, as it is the smallest model among the
three considered architectures. In both methods, the first step is to perform KD to reduce the number of parameters for computational efficiency during inference while learning better representations for AMC. To achieve this, the pre-trained InceptionNet is used as the teacher model for the VTCNN2 student model (i.e., case \Romannum{1} in Table \ref{tab:kd_cases}). The training parameters are the same as the ones used while evaluating the KD method. For the DP method, we then apply the NT pruning on the first FC layer of the distilled VTCNN2 with $\epsilon=0.08$, as it achieves maximum sparsity for VTCNN2 (96.5\%). For the DQ method, the PQ algorithm is performed on the first FC layer of the distilled VTCNN2, and $P=2$ is used as it provides the highest compression rate for VTCNN2 ($39.65$). It can be observed in Fig. \ref{fig:dp-dq} that both the DP and DQ optimized models can achieve marginally better classification accuracies than the benchmark. Table \ref{tab:all_comp} compares the individual and combined strategies for VTCNN2 in terms of accuracies and optimization factors for each method. It is noticed that the DQ method can maintainin a similar compression rate ($C_Q = 39.65$) to the individual PQ method with $P=2$ with slight improvement in the accuracy. With the DP method, we can achieve a higher $p_e$ of $97.1$\% compared to the NT method, which achieves $96.5$\% with $\epsilon=0.08$ for VTCNN2.

Thus, it can be established that the proposed individual and combined strategies are highly effective for optimizing DL-based AMC models while maintaining or improving classification performance. Once optimized for performance predeployment, the models can achieve higher storage benefits and faster inference on resource-constrained edge devices. Also,
for large models with high complexity and high performance, 
a combination of optimization methods, like the proposed DP and DQ methods can be more beneficial in developing smaller, reduced parameter models while merging the complementary benefits of sparsity or compression to provide better-optimized models. The use of the combined methods based on the use case (i.e., whether to sparsify or to quantize) can be an excellent possibility for edge applications of AMC.
\vspace{-0.1em}
\section{Conclusions}
\vspace{-0.1em}
This work showed the effectiveness  of three individual model optimization methods: the NT algorithm, the PQ algorithm, and the vanilla KD on CNN models developed for AMC. These methods can optimize the models based on sparsity, compression, and parameter reduction, without substantial loss in classification accuracy and can even improve the performances  using KD. This is advantageous for deploying accurate AMC models on edge devices. The proposed optimized models using the NT algorithm could achieve sparsity as high as $98\%$, and the PQ algorithm achieved a compression rate of 133 with minimal loss in accuracy. We observed that combination strategies: the DP and DQ methods, can effectively optimize a model, with the DP method having even higher sparsity ($97.1\%$) for VTCNN2 compared to the NT method ($96.5\%$). Our future work will involve developing optimization techniques tailored to other desired optimization criteria for edge applications. 
\begin{table}[t!]
    \centering
    \caption{\small Comparison of proposed individual and combined strategies of model optimization for VTCNN2. \vspace*{-0.3em}}
    \renewcommand{\arraystretch}{1.1}
    
    \setlength{\tabcolsep}{5pt}
    \begin{tabular}{|M{0.75cm}|M{0.6cm}|M{0.6cm}|M{1.3cm}|M{0.75cm}|M{0.7cm}|M{1.3cm}|}
        \hline
           \multirow{2}{*}{Method} & \multirow{2}{*}{$p_e$} & \multirow{2}{*}{$C_Q$} & \multirow{2}{*}{Accuracy} & \multirow{2}{*}{Sparsity} & \multicolumn{2} { | c | }{Efficiency} \\ \cline{6-7}
          & &  &  & & Storage & Computation\\
    \hline
        \small NT  & 96.5\% & - &  Comparable & \checkmark & \checkmark & \checkmark\\ 
    \hline
         PQ  & - & 39.65 & Comparable & - & \checkmark & -\\ 
    \hline
         KD  & - & - & Improved & - & \checkmark & \checkmark\\
    \hline
       DQ & - & 39.65 & Marginally improved & - & \checkmark & \checkmark\\
    \hline
        DP & 97.1\% & - &  Marginally improved & \checkmark & \checkmark & \checkmark\\
    \hline
    \end{tabular}
    \label{tab:all_comp}
\end{table}
\vspace*{-1em}
\bibliographystyle{IEEEtran}
\bibliography{references}

\begin{thebibliography}{10}
\providecommand{\url}[1]{#1}
\csname url@samestyle\endcsname
\providecommand{\newblock}{\relax}
\providecommand{\bibinfo}[2]{#2}
\providecommand{\BIBentrySTDinterwordspacing}{\spaceskip=0pt\relax}
\providecommand{\BIBentryALTinterwordstretchfactor}{4}
\providecommand{\BIBentryALTinterwordspacing}{\spaceskip=\fontdimen2\font plus
\BIBentryALTinterwordstretchfactor\fontdimen3\font minus
  \fontdimen4\font\relax}
\providecommand{\BIBforeignlanguage}[2]{{%
\expandafter\ifx\csname l@#1\endcsname\relax
\typeout{** WARNING: IEEEtran.bst: No hyphenation pattern has been}%
\typeout{** loaded for the language `#1'. Using the pattern for}%
\typeout{** the default language instead.}%
\else
\language=\csname l@#1\endcsname
\fi
#2}}
\providecommand{\BIBdecl}{\relax}
\BIBdecl

\bibitem{nandi}
E.~E. Azzouz and A.~K. Nandi, ``Automatic modulation recognition of
  communication signals,'' \emph{IEEE Trans. Commun.}, vol. 334, no.~4, pp.
  431--436, 1998.

\bibitem{dobre2007survey}
O.~A. Dobre, A.~Abdi, Y.~Bar-Ness, and W.~Su, ``Survey of automatic modulation
  classification techniques: {C}lassical approaches and new trends,'' \emph{IET
  commun.}, vol.~1, no.~2, pp. 137--156, Apr. 2007.

\bibitem{o2016convolutional}
T.~J. O’Shea, J.~Corgan, and T.~C. Clancy, ``Convolutional radio modulation
  recognition networks,'' in \emph{Proc. Int. Conf. Eng. Appl. Neural Netw.
  (EANN)}, Aberdeen, UK, Sep. 2--5, 2016, pp. 213--226.

\bibitem{west2017deep}
N.~E. West and T.~O'shea, ``Deep architectures for modulation recognition,'' in
  \emph{Proc. IEEE Int. Symp. Dyn. Spectr. Access Netw. (DySPAN)}, Baltimore,
  MD, USA, Mar. 6--9, 2017, pp. 1--6.

\bibitem{liu2017deep}
X.~Liu, D.~Yang, and A.~El~Gamal, ``Deep neural network architectures for
  modulation classification,'' in \emph{Proc. Asilomar Conf. Signals, Syst.,
  Comput.}, Pacific Grove, CA, USA, Oct. 29--Nov. 1, 2017, pp. 915--919.

\bibitem{zhang2020automatic}
Z.~Zhang, H.~Luo, C.~Wang, C.~Gan, and Y.~Xiang, ``Automatic modulation
  classification using {CNN-LSTM} based dual-stream structure,'' \emph{IEEE
  Trans. Veh. Tech.}, vol.~69, no.~11, pp. 13\,521--13\,531, Nov. 2020.

\bibitem{mishra2020survey}
\BIBentryALTinterwordspacing
R.~Mishra, H.~P. Gupta, and T.~Dutta, ``A survey on deep neural network
  compression: Challenges, overview, and solutions,'' Oct. 2020. [Online].
  Available: \url{https://arxiv.org/abs/2010.03954}
\BIBentrySTDinterwordspacing

\bibitem{cheng2017survey}
\BIBentryALTinterwordspacing
Y.~Cheng, D.~Wang, P.~Zhou, and T.~Zhang, ``A survey of model compression and
  acceleration for deep neural networks,'' Oct. 2017. [Online]. Available:
  \url{https://arxiv.org/abs/1710.09282}
\BIBentrySTDinterwordspacing

\bibitem{shi2020communication}
Y.~Shi, K.~Yang, T.~Jiang, J.~Zhang, and K.~B. Letaief,
  ``Communication-efficient edge {AI}: Algorithms and systems,'' \emph{IEEE
  Commun. Surveys \& Tuts.}, vol.~22, no.~4, pp. 2167--2191, Jul. 2020.

\bibitem{aghasi2017net}
A.~Aghasi, A.~Abdi, N.~Nguyen, and J.~Romberg, ``Net-trim: Convex pruning of
  deep neural networks with performance guarantee,'' in \emph{Proc. Adv.
  Neural. Inf. Process. Syst. (NIPS)}, Long Beach, CA, USA, Dec. 4--7, 2017,
  pp. 3180--3189.

\bibitem{denil2013predicting}
M.~Denil, B.~Shakibi, L.~Dinh, M.~A. Ranzato, and N.~De~Freitas, ``Predicting
  parameters in deep learning,'' in \emph{Proc. Adv. Neural Inf. Process. Syst.
  (NIPS)}, Lake Tahoe, NV, USA, Dec. 5--10, 2013, pp. 2148--2156.

\bibitem{jegou2010product}
H.~Jegou, M.~Douze, and C.~Schmid, ``Product quantization for nearest neighbor
  search,'' \emph{IEEE Trans. Pattern Anal. Mach. Intell.}, vol.~32, no.~1, pp.
  1--15, Jan. 2010.

\bibitem{hinton2015distilling}
\BIBentryALTinterwordspacing
G.~Hinton, O.~Vinyals, and J.~Dean, ``Distilling the knowledge in a neural
  network,'' Mar. 2015. [Online]. Available:
  \url{https://arxiv.org/abs/1503.02531}
\BIBentrySTDinterwordspacing

\bibitem{vtcnn2}
T.~O'Shea and N.~West, ``Radio machine learning dataset generation with {GNU}
  {R}adio,'' in \emph{Proc. GNU Radio Conf.}, Boulder, CO, USA, Sep. 2016.

\bibitem{yu2017scalpel}
J.~Yu \emph{et~al.}, ``Scalpel: Customizing dnn pruning to the underlying
  hardware parallelism,'' in \emph{Proc. IEEE/ACM 44th Annu. Int. Symp. Comput.
  Architecture (ISCA)}, Toronto, ON, Canada, Jun. 24--28, 2017, pp. 548--560.

\bibitem{zhang2020snap}
J.~F. Zhang \emph{et~al.}, ``Snap: An efficient sparse neural acceleration
  processor for unstructured sparse deep neural network inference,'' \emph{IEEE
  J. Solid-State Circuits}, vol.~56, no.~2, pp. 636--647, Feb. 2021.

\end{thebibliography}
\end{document}